\documentstyle[twoside,fleqn,epsf,espcrc2]{article}
 
\newcommand{\AmS}{{\protect\the\textfont2
  A\kern-.1667em\lower.5ex\hbox{M}\kern-.125emS}}
 
\title{Hybrid Monte Carlo Without Pseudofermions}
 
\author{K. F. Liu, S. J. Dong and C. Thron\address{Department of Physics
and Astronomy, University of Kentucky, Lexington, KY 40506, USA}
 \thanks{This work is partially supported by DOE grant
        No. DE-Fg05-84ER40154.}}

\begin{document}
 
\begin{abstract}
We introduce a dynamical fermion algorithm which is based on the
hybrid Monte Carlo (HMC) algorithm, but without pseudofermions.
The molecular dynamics steps in HMC are retained except the derivatives
with respect to the gauge fields are calculated with the $Z_2$ noise.
The determinant ratios are estimated with the Pa\`{d}e - $Z_2$
method. Finally, we use the Kennedy-Kuti linear accept/reject method for
the Monte Carlo step which is shown to respect detailed balance.
We comment on the comparison of this algorithm with the pseudofermion
algorithm.
 
\end{abstract}
 
\maketitle
 
\section{Introduction}
 
The partition function of the gauge field theory with dynamical fermions
in the Euclidean path-integral formalism has the following generic form
 
\begin{equation} \label{partition}
Z = \int [dU] det M[U] e^{- S_G}
\end{equation}
where $S_G$ is the pure gauge part of the action and U is the link
variable. The Grassmann numbers representing the fermions are formally
intergrated out to give a fermion determinant. Since the determinant is
very non-local, the Monte Carlo algorithms with dynamical fermions is
known to be very time-consuming. The best algorithm so far is the
Hybrid Monte Carlo \cite{dkp87} which introduces a Monte Carlo step to
the hybrid (molecular/hybrid) algorithm to remove the discretization
error from the molecular dynamics step. This algorithm avoids calculating
the determinant directly by introducing pseudofermion variables. The
partition then takes the form
 
\begin{equation}
Z = \int [dU][d\chi^*][d\chi] e^{-S_G -\chi^*(M^{\dagger}M)^{-1}\chi}
\end{equation}
where $\chi$ is the pseudofermion field variable.
 
  In this paper, we shall present an algorithm which estimates the
determinant ratio directly without using pseudofermions. To begin with,
we note that the determinant can be written as $e^{Tr\,\log M}$ and the
trace can be estimated by the $Z_2$ noise \cite{dl94} which is known
to yield the minimum variance \cite{bmt94}. Thus the partition in
eq. (\ref{partition}) takes the form
 
\begin{equation}  \label{partition-z}
Z = \int [dU]e^{-S_G + \int dz_2\, z_2 (\log M) z_2}
\end{equation}
where $z_2$ is the $Z_2$ noise for estimating the trace of the matrix
$\log M$.
 
\section{Estimating Determinants with Pad\`{e}-$Z_2$ Method}
 
It has been shown that there is an efficient
algorithm to estimate the determinant by expanding the  $\log M$
using the Pad\`{e} approximation. The trace is then estimated by the
$Z_2$ noise. This is reported by Chris Thron \cite{thr96}. The
expectation value of the $\log det M$ can be written as
 
\begin{equation}
E[\log det M] = E[Tr \log M] %= \frac{1}{N} \sum_{j} z_2^j \log M z_2^j
\sim \frac{1}{N}  \sum_{i,j} z_2^j (\frac{b_i}{M + c_i I}) z_2^j
\end{equation}
where $b_i$ and $c_i$ are constants. It is worthwhile noting that
since we will use this estimate for the determinant ratios in the
Monte Carlo step for the two matrices $M_1$ and $M_2$ whose eigenvalues
are close to each other, we can do the Pad\`{e} expansion aroud 1
with a small set of $b_i$ and $c_i$. As long as there is enough
memory, the matrix inversion with the set of $b_i$ and $c_i$ can be
done with the minimum residual with about 8\% overhead in CPU time
as compared with inversion of the one with the smallest diagonal term
(the equivalent of the quark mass) \cite{fhn94,ggl96,ysl96}.
Furthermore, $c_i$ are real
and positive so that the matrix $M + c_i I$ is better conditioned
then M itself. Finally, we should caution that the PZ method does not
work if there are real negative eigenvalues, because the $\log$ has a
cut on the negative real axis.
 
\section{Hybrid Monte Carlo with PZ Determinant}
 
   The algorithm we propose involves two steps:
 
\noindent
1. {\bf Molecular Dynamics with Determinant Force:}
 
The molecular dynamics step is retained as in HMC, except the
pseudofermion force is replaced with the determinant force in the
evolution equations.
 
\begin{flushleft}
\begin{eqnarray}
\pi(\frac{\delta\tau}{2})&\!\!\! =\!\!\!& \pi(0) - T[\frac{\partial S_G}
{\partial U}(0)U(0)]\frac{\delta\tau}{2} \nonumber \\
 &\!\!\!+\!\!\!& z_2 T[\frac{\partial \log M}
{\partial U}(0)U(0)]z_2\frac{\delta\tau}{2} \nonumber \\
U(\delta\tau) &\!\!\! =\!\!\!& exp[\pi(\frac{\delta\tau}{2})\delta\tau]
 U(0) \nonumber \\
\pi(\delta\tau) &\!\!\! =\!\!\!& \pi(\frac{\delta\tau}{2}) -
T[\frac{\partial S_G}{\partial U}(\delta\tau)U(\delta\tau)]
\frac{\delta\tau}{2} \nonumber\\
&\!\!\!+\!\!\!& z_2 T[\frac{\partial \log M}
{\partial U}(\delta\tau)U(\delta\tau)]z_2\frac{\delta\tau}{2}
\end{eqnarray}
\end{flushleft}
 
Several comments are ready:
 
The leapfrog scheme is adopted for this step as in HMC to make it
reversible and area preserving.
There is a discretization error $N_{MD} O(\delta\tau^3) =
O(\delta\tau^2)$ as in HMC.
Like the pseudofermion variable in HMC, $z_2$ is kept fixed during the
MD trajectory.
 
\noindent
2. {\bf Monte Carlo:}
 
In lieu of the Metropolis accept/reject criteria, we shall us the
Kennedy-Kuti linear test \cite{kk85}
 
$P(V \longleftarrow U) = \left\{ \begin{array}{ll}
   \frac{1}{2}  &  \mbox{if $U > V$} \\
           \langle\frac{R}{2}\rangle_{z_2} & \mbox{if $U \leq V$}
\end{array}
\right. $  \\
where U and V are two configurations and $R = e^{-\delta H}$.
H is molecular dynamics hamiltonian. We note that
this satisfies detailed balance provided $R$ is an unbiased estimate
\cite{kk85}.
Since the estimate of $\delta H$ is unbiased with the $Z_2$ noise,
we can use the Bhanot-Kennedy algorithm \cite{bk85} to turn $R$ into
an unbiased estimate: \\
Suppose x is an unbiased estimate,
\begin{eqnarray}  \label{bk}
e^x &=& 1 + x + \frac{1}{2}(x^2 + \frac{1}{3}(x^3 + ... \nonumber \\
\langle e^{\hat{x}}\rangle &=& e^x
\end{eqnarray}

\section{Comparison with HMC with Pseudofermions}
We shall point out several apparent differences between the
pseudofermion algorithm and the current determinant algorithm and
will make some preliminary comparison of the two approaches.
 
\noindent
1. {\bf MD trajectories:}
 
The pseudofermion approach requires two matrix inversions
$\longrightarrow M^{\dagger}M x = \chi$ in each step. Whereas, the
determinant approach reqires only one such inversion in
$\longrightarrow M x = z_2$.
Accuracy of the residual of matrix inversion is needed for reversibility
for both approaches, i.e.
\begin{equation}
\langle e^{- \delta H}\rangle = 1
\end{equation}
where the average is over all the trajectories. This allows one to
test the thermal equilibrium and make a comparison of the two approaches.
 
\noindent
2. {\bf Overhead of Determinant Algorithm in Accept/Reject Step:}
 
The Bhanot-Kennedy unbiased estimate in eq. (\ref{bk}),
needs on the average e terms in the matrix inversion to have
independent estimate of $\delta H$. It may be desirable to have more
than one $Z_2$ noise in the estimate of the determinant ratio.
 
In the Kennedy-Kuti linear accept/reject test, the ratio $R$ should
lie between 0 and 2 to be a probability. This, however, can be violated
in practice \cite{bk85}. To make sure that the
systematic errors due to the violation of $0 < R < 2$
is small, say less than 0.1\%, one needs to have small enough step
size in the MD trajectories.
 
\noindent
3. {\bf Intrinsic Noisiness:}
 
The ultimate test between different algorithms will be conducted by
comparing how fast observables decorrelate. This will be done next. In the
meantime, we shall concentrate on the issue of intrinsic noisiness.
While the pseudofermion method gives the joint probability of gauge
fields and pseudofermion fields, the determinant method has built in
it the fluctuation of the $Z_2$ noise. We shall explore the intrinsic
noisiness by computing $det M_1/ det M_2$ for a fixed pair of
$M_1$ and $M_2$ separated by heatbath update of 20 links on a $16^3
\times 24$ lattice with $\kappa = 0.154$ with the two different methods.
 
In the pseudofermion method, the determinant ratio is estimated by
\begin{equation}
det M_1/ det M_2 \sim \frac{1}{N} \sum_{i=1}^N
e^{-\chi_i^* (M_2^{\dagger} M_2)^{-1} \chi_i - \eta_i^2}
\end{equation}
where $\chi_i = M_1^{\dagger} \eta_i$.
In the PZ method , the determinant ratio is estimated by
\begin{equation}
log (\frac{det M_1}{ det M_2}) \sim \frac{1}{N} \sum_{i=1}^N
\sum_{j=1}^L z_2^j(\frac{b_i}{M_1/M_2 + c_i})\,z_2^j
\end{equation}
 
We plot in Fig. 1 and Fig. 2 the distribution of $\log (det M_1/ det M_2)
- \langle\log (det M_1/ det M_2)\rangle$ for 50 gaussian and $Z_2$
noises respectively. We see that the distribution of the pseudofermion
method is wider than that of the PZ method. We also plot in Fig. 3 the
Jackknife errors of $\log (det M_1/ det M_2)$ due to the two method as
a function of the number of noise. Again we see that the PZ method yields
smaller error than that of the pseudofermion method.
 
 Extensive tests are needed to check the efficiency of this algorithm
before we can tell if this algorithm will turn out to a useful one.

\begin{figure}[ht]
\vspace{9pt}
\setlength\epsfxsize{70mm}
\epsfbox{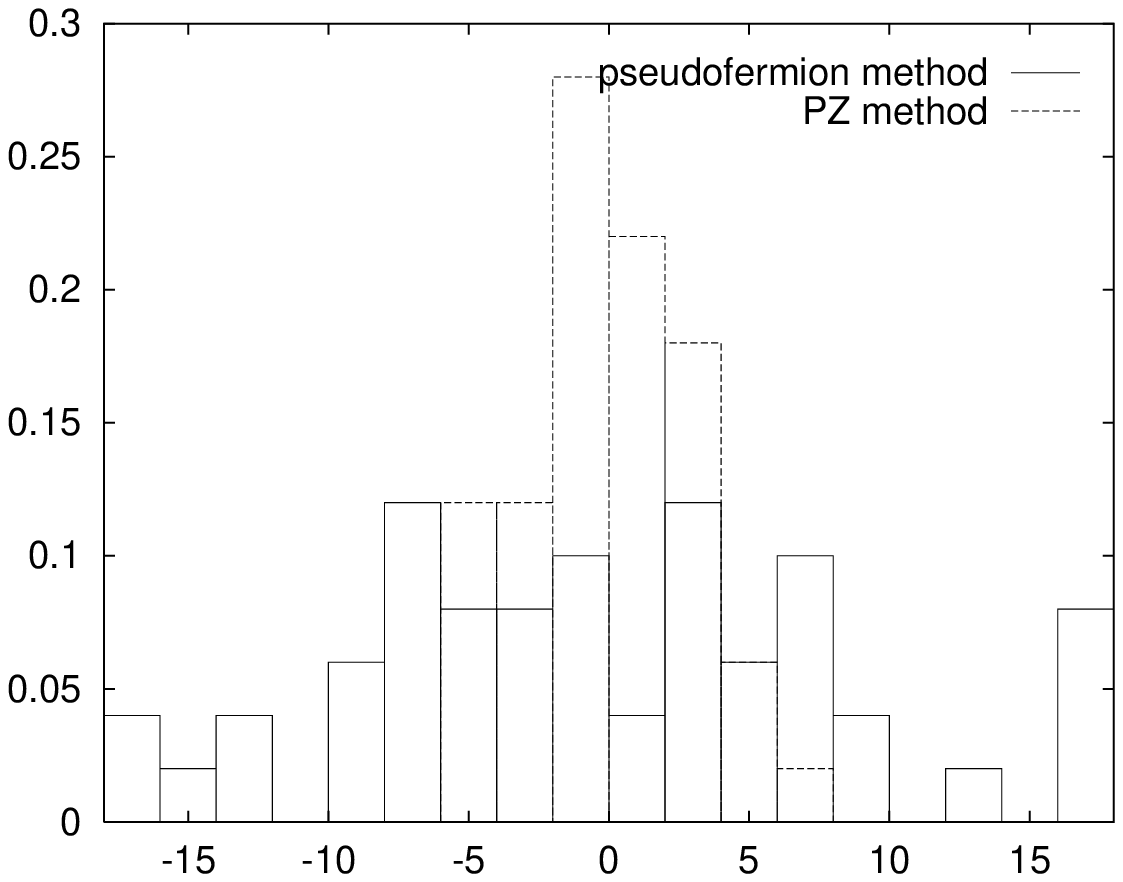}
%%\framebox[55mm]{\rule[-21mm]{0mm}{43mm}}
\caption{distributions of $\log (det M_1/ det M_2)
- \langle\log (det M_1/ det M_2)\rangle$ dor gaussian and $Z_2$ noises}
%%\label{fig:largenenough}
%%\end{figure}
%
%%\begin{figure}[ht]
%%\framebox[55mm]{\rule[-21mm]{0mm}{43mm}}
\setlength\epsfxsize{70mm}
\epsfbox{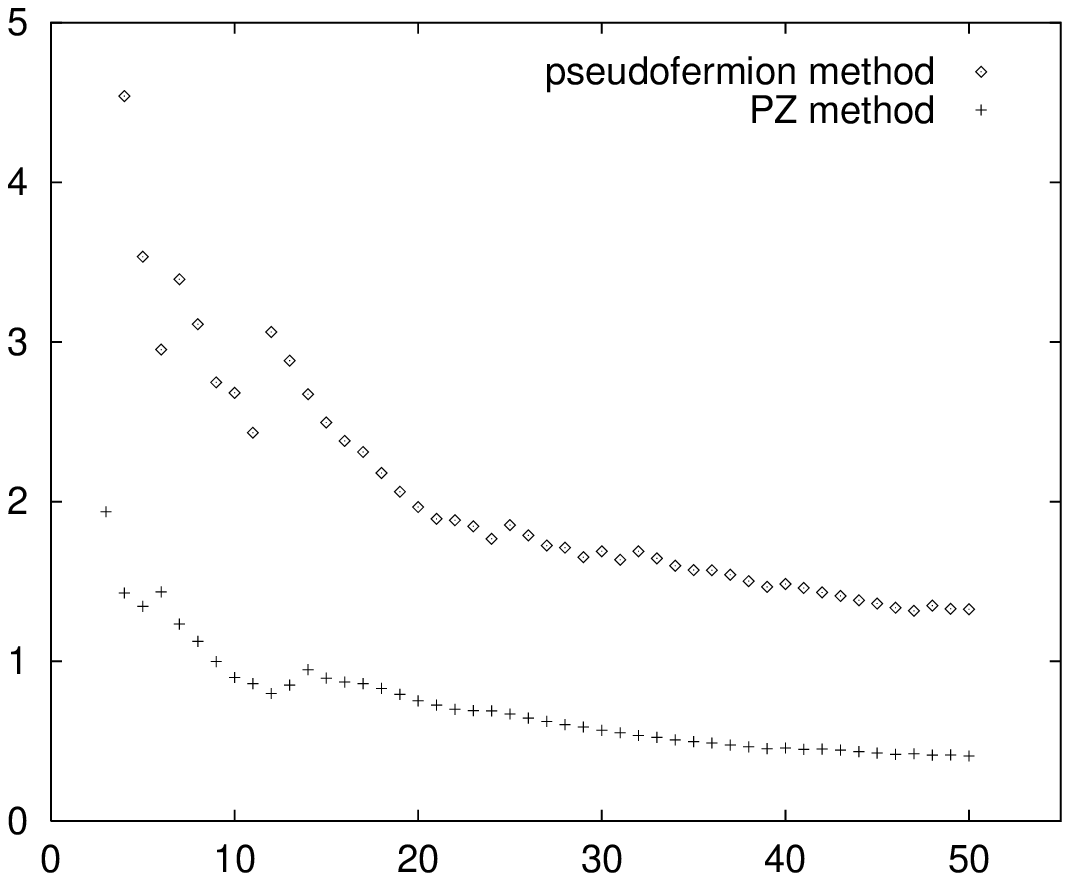}
\caption{Jackknife errors of $\log (det M_1/ det M_2)$}
%%\label{fig:toosmall}
\end{figure}

\end{document}